\documentclass[%
 reprint,
nofootinbib,
 amsmath,amssymb,
 aps,
 prd,
 floatfix,
]{revtex4-2}
\usepackage{graphicx}
\usepackage{dcolumn}
\usepackage{bm}

\usepackage{tikz}
\usepackage{booktabs}
\usepackage{array}
\usepackage{multirow}
\usepackage{mathtools}
\usepackage{cancel}
\usepackage{subfigure}
\usepackage{comment}
\usepackage{float}
\usepackage[normalem]{ulem}
\usepackage{graphicx}
\usepackage{diagbox}
\usepackage{txfonts}

\begin{document}

\title{GRMHD simulations of accretion flows onto unequal-mass,\\ precessing massive binary black hole mergers}
\author{Federico Cattorini$^{1,3}$}
 \email{cattorini.federico@gmail.com}
\author{Bruno Giacomazzo$^{1,3,4}$}
\author{Monica Colpi$^{1,3,4}$}
\author{Francesco Haardt$^{2,3,4}$}
\affiliation{%
$^1$Dipartimento di Fisica G. Occhialini, Universit\`a di Milano-Bicocca, Piazza della Scienza 3, I-20126 Milano, Italy}%
\affiliation{$^2$Dipartimento di Scienza e Alta Tecnologia, Universit\`a degli Studi dell'Insubria, Via Valleggio 11, I-22100, Como, Italy}
\affiliation{%
 $^3$INFN, Sezione di Milano-Bicocca, Piazza della Scienza 3, I-20126 Milano, Italy}%
 \affiliation{%
 $^4$INAF, Osservatorio Astronomico di Brera, Via E. Bianchi 46, I-23807 Merate, Italy}%

\date{\today}

\begin{abstract}
In this work, we use general relativistic magnetohydrodynamics simulations to explore the effect of spin orientation on the dynamics of gas in the vicinity of merging black holes. We present a suite of eight simulations of unequal-mass, spinning black hole binaries embedded in magnetized clouds of matter. Each binary evolution covers approximately 15 orbits before the coalescence. The geometry of the accretion flows in the vicinity of the black holes is significantly altered by the orientation of the individual spins with respect to the orbital angular momentum, with the primary black hole dominating the mass accretion rate $\dot{M}$. We observe quasiperiodic modulations of  $\dot{M}$ in most of the configurations, whose amplitude is dependent on the orientation of the black hole spins. 
We find the presence of a relation between the average amplitude of $\dot{M}$ and the spin precession parameter $\chi_{\mathrm{p}}$ showing  that spin misalignment systematically leads to stronger modulation, whereas configurations with spins aligned to the orbital angular momentum damp out the quasiperiodicity. This finding suggests a possible signature imprinted in the accretion luminosity of precessing binaries approaching merger and has possible consequences on future multimessenger observations of massive binary black hole systems.
\end{abstract}

\pacs{
04.25.D-	
04.30.Db	
95.30.Qd	
97.60.Lf	
}
\maketitle

\section{Introduction}
Gravitational waves (GWs) generated by binaries of MBHs in the mass range $M$$\sim$$10^4$\textendash$10^7 \ \mathrm{M}_{\odot}$ are expected to be detected by future space-based, low-frequency gravitational interferometers such as the Laser Interferometer Space Antenna \cite[LISA,][]{LISA,LISA-2017, LISA2023} and the Taiji Program in Space \cite[]{Taiji}. 
In the aftermath of a gas-rich galaxy merger, the late-inspiral and coalescence of MBBHs is expected to occur in gas-rich environments \cite[][]{Kocsis-2006, Capelo-2015, LISA2023}, making these sources particularly promising systems for multimessenger astrophysics.
Concurrent (multimessenger) detection of gravitational and electromagnetic (EM) radiation generated by merging MBBHs yields the potential to resolve a number of astrophysical conundrums, e.g., providing new probes of the universe expansion \cite[][]{Schutz-2018, Abbott-2017}, and offering new opportunities to deepen our understanding of MBHs evolution in the context of large-scale structures \cite[][]{Kormendy-2013}. 
Future observational endeavors searching for these systems rely on firm predictions for the EM emission arising before, during, and after the merger. Without accurate predictions, it will be challenging to know what to look for and where to search in the EM spectrum. Limited knowledge of the unique spectral and timing features of such signals would frustrate future multimessenger efforts. Thus, next-generation multimessenger astronomy relies on detailed theoretical understanding of the mechanisms that may give rise to EM signatures of MBBHs, aiming to determine distinctive features that will help distinguishing them from other events.

There are several proposed models about the gaseous environment in which a MBBH can be embedded. When the gas around the binary has enough angular momentum, the two MBHs are expected to be surrounded by a dense and cold, radiatively inefficient \textit{circumbinary disk} (CBD) with the black holes located in a central low-density region excavated by the binary tidal torques \cite[][]{Milosavljevic-2005, Kocsis-2012MNRAS.427.2680K, MacFadyen-2008}. The CBD feeds mass to the black holes through tidal streams, which may form individual accretion disks (``mini-disks'') around each MBH \cite[][]{Bowen-2018, Bowen-2019, Combi-2022, Bright-2023}. Over the last decade, this scenario has been extensively studied by several theoretical groups adopting a variety of different numerical techniques \cite[see, e.g.,][for a review]{Gold-2019}.

Since the balance between heating and cooling mechanisms in the gas near MBBHs can be significantly altered toward the merger, different scenarios for the accretion flow in the proximity of merging MBBHs are possible. If radiative cooling is inefficient, the gaseous environment would resemble a hot and tenuous radiatively inefficient accretion flow \cite[RIAF, ][]{Ichimaru-1977, Narayan-1994}. In this scenario, most of the energy generated by accretion and turbulent stresses is stored in the gas, resulting in a geometrically thick accretion flow \cite[][]{Bogdanovic-2011}. Furthermore, the binary tidal torques are incapable of creating a central cavity, because the ejected gas is replenished on a dynamical time scale. Therefore, the binary will find itself engulfed in a hot \textit{gas cloud} all the way down to the merger.

The gas cloud scenario has been modeled in relativistic simulations by several groups \cite[][]{Palenzuela-2010, Bode-2010, Giacomazzo-2012, Kelly-2017}. In \cite{Cattorini-2021}, we have explored the features of moderately magnetized gas cloud accretion onto equal-mass, aligned-spinning binaries by producing a set of nine ideal-GRMHD simulations covering a range of initially-uniform magnetized fluids with different initial magnetic-to-gas pressure ratio $\beta_0^{-1}$. Our results have shown that aligned-spins have a suppressing effect on the mass accretion rate as large as $\sim$50\%; also, we found that the peak Poynting luminosity reached shortly after merger is enhanced by up to a factor $\sim$2.5 for binaries of spinning black holes compared to nonspinning configurations.
In the follow-up investigation \cite{Cattorini-2022}, we produced five simulations of equal-mass binaries of spinning black holes characterized by diverse orientations of the spins relative to the orbital angular momentum.
Notably, we  found that the orientation of the individual spins during the late-inspiral appears to be related with quasiperiodic modulations in the rate of mass accretion, leading to a ``chirp'' in the accretion rate that strongly resembles the gravitational one.

In this Paper, we present a suite of eight unequal-mass simulations of binary black holes immersed in a cloud of magnetized matter and consider a broader family of spin-misalignment. We aim to further explore the relation between the orientation of the individual spins and the quasiperiodic modulations in the mass accretion rate.
The Paper is organized as follows. In Section \ref{sec:numerical}, we outline the numerical methods adopted in our simulations and provide the initial data for the metric and the MHD fields. Section \ref{sec:results} is devoted to our results on the effects of spin-misalignment on the dynamics of magnetized accretion flows and the link between spin-orientation and quasiperiodicities. Finally, in Section \ref{sec:con} we summarize our conclusions.
Throughout the Paper, we adopt geometric units and set $c=G=M=1$, where $M$ is the total mass of the system.
\section{Numerical methods and initial data}\label{sec:numerical}
In the present Section, we outline the numerical setup adopted to solve numerically Einstein's field equations and the equations of ideal MHD in curved and dynamical spacetime.
All our simulations have been carried out on adaptive-mesh refinement (AMR) grids provided by the Carpet driver \cite[][]{Schnetter-2004} using the \texttt{Einstein Toolkit}\footnote{\url{http://einsteintoolkit.org}} framework \cite[][]{Loffler-2012, Zilhao-2013, ETzenodo}, a powerful infrastructure for relativistic astrophysics and gravitational physics made up by several components (the so-called ``thorns'').

We produced a suite of eight simulations of unequal-mass binaries (Table \ref{tab:initdata}) starting on quasicircular orbits at a orbital coordinate separation $a_0 = 12.5 M$ and evolving across the late-inspiral and merger. 
Along the path to coalescence of a MBBH, the accretion of matter onto the binary components tends to drive their mass-ratio close to unity \cite[][]{Farris-2014a, Duffell-2020}. For this reason, we have chosen to evolve equal-mass systems and unequal-mass systems with mass ratios close to unity. For all our models, the binary mass-ratio is $q \equiv m_-/m_+ = 0.6$, where $m_-$ ($m_+$) is the mass of the secondary (primary) black hole. 
We produced eight $q=0.6$ simulations of BBHs with spin magnitude $a=0.6$ and covering a broader range of spin misalignments with the orbital angular momentum.

Six configurations consider binaries immersed in a magnetized plasma with an adiabatic index $\Gamma=4/3$. In addition, we evolve two models with the same initial metric setup of configuration \texttt{b0p6} and with an adiabatic index $\Gamma=5/3,13/9$. 

The fluid and magnetic-field configurations are equivalent to those by \cite{Cattorini-2022}: the binary systems are immersed in an initially uniform, radiation-dominated polytropic fluid ($p_0 = \kappa \rho_0^{\Gamma}$, with $\rho_0 = 1$, \ $\kappa = 0.2, \ \Gamma = 4/3$). 
The fluid is initially at rest relative to black holes, which is non-physical. Therefore, we must be careful to start our BBH at a large enough separation to allow the gas in the strong-field region to establish a quasi-equilibrium flow with the binary motion. The numerical explorations by \cite{Kelly-2017} have investigated the dependence of the timing features of the evolving plasma on the initial binary separation and observed that binary configurations starting at initial separations of $11.5 M$, $14.4 M$, and $16.3 M$ show the same qualitative behavior. Motivated by this result, we choose to evolve our binaries starting from an initial separation $12.5 M$.
We evolve our binaries over the last $\sim$15 orbits to merger and beyond.

The fluid in which the MBBHs are embedded is threaded by an initially uniform magnetic field parallel to the binary angular momentum, i.e., $B^i=(0,0,B^z)$. The magnetic field is assumed to be anchored to a distant circumbinary disk located outside the computational domain. This initial configuration of the magnetic field is analogous to that implemented in previous works \cite[e.g.,][]{Palenzuela-2010, Palenzuela-2010b, Moesta-2012, Giacomazzo-2012, Kelly-2017}.
As in \cite{Cattorini-2022}, we choose the initial magnetic-to-gas pressure to be $\beta_0^{-1}=0.31$.

Our simulations adopt a cubic domain given by $[-1024M, 1024M]^3$ and employ AMR with $N=11$ levels of refinement. The coarsest resolution is $\Delta x_c = 128M/7$, and the finest one is $\Delta x_f = \Delta x_c \cdot 2^{1-N} = M/56$. We employ radiative (or Sommerfeld) boundary conditions for the metric and “outflow" boundary conditions \cite[][]{Etienne-2015} to the MHD variables.
\begin{table*}
\centering
\caption{BBH initial data parameters in code units of the GRMHD runs: initial puncture separation $a_0$, individual puncture masses $m_+/M$ and $m_-/M$,  linear momentum components $p_r$ \& $p_t$, and dimensionless spin vectors $\hat{a}_i = (a_{i,x}, a_{i,y}, a_{i, z})$ of each BH. Data for model \texttt{b0p6} refer to three different simulations, characterized by different values of the adiabatic index $\Gamma$.}\label{tab:initdata}
 \begin{tabular}{lcccc}
\hline\hline \\[-1.6ex]
  Run & \ $|p_{r}|$ \ & \ $|p_{t}|$  \ \ &  \ \ $\hat{a}_{+}$ \ $[m_+^2]$ \ \ & \ \ $\hat{a}_{-}$ \ $[m_-^2] $ \ \\
  \cmidrule(lr) {1-5}
  \texttt{b0p2}  & \ \  3.61e-4  \ \ &  7.53e-2 & \ \  \ \ (0.00, \ 0.00,  \ 0.60)  \ \ &  (0.00, \ 0.00, \ 0.60)  \ \  \\
\cmidrule(lr) {1-5}
 \texttt{b05p12}   & \ \  3.62e-4 \ \  &  7.53e-2 &  \ \ (-0.16, \ 0.00, \ 0.58) &  (0.16,  \ 0.00, \ 0.58)  \ \  \\
 \cmidrule(lr) {1-5}
  \texttt{b0p3}   & \ \  3.67e-4  \ \  &  7.56e-2 &  \ \ (-0.30, \ 0.00, \ 0.52) & (0.30, \ 0.00,  \ 0.52)  \ \  \\
\cmidrule(lr) {1-5}
  \texttt{b0p4}  &  \ \ 3.74e-4  \ \ &  7.59e-2 &  \ \ \ \  (-0.42, \ 0.00, \ 0.42)  \ \ & (0.42, \ 0.00,  \ 0.42)  \ \    \\
 \cmidrule(lr) {1-5}
  \texttt{b0p6}   &  \ \ 3.85e-4  \ \ & 7.64e-2 &  \ \  \ \ (-0.52, \ 0.00, \ 0.30)  \ \ & (0.52, \ 0.00, \ 0.30)  \ \  \\
\cmidrule(lr) {1-5}
  \texttt{b0p12} & \ \  3.99e-4 \ \  &  7.70e-2 &  \ \  \ \ (-0.58, \ 0.00, \ 0.16) \ \  & (0.58, \ 0.00, \ 0.16)  \ \ \\
\hline
\hline
 \end{tabular}
\end{table*}
\subsection{Spacetime metric evolution}\label{sec:metric_evo}
We write the generic line element in the standard 3+1 form as:
\begin{equation}\label{eq:line3+1}
 d s^{2}=g_{\mu \nu} d x^{\mu} d x^{\nu}=-\alpha^{2} d t^{2}+\gamma_{i j}\left(d x^{i}+\beta^{i} d t\right)\left(d x^{j}+\beta^{j} d t\right),
\end{equation}
where $\alpha$, $\beta^i$, and $\gamma_{ij}$ are the lapse function, shift, and spatial metric, respectively. 
We evolve $\gamma_{ij}$ and the extrinsic curvature $K_{ij}$ using the Kranc-based \texttt{McLachlan} \cite[][]{Husa-2006, Brown-2009} code, which evolves Einstein's equations adopting the BSSN \cite[][]{Nakamura-1987, Shibata-Nakamura-1995, Baumgarte-Shapiro-1998} formalism \cite[the evolution and constraint equations for the spacetime fields in the BSSN form are summarized, e.g., in][]{baumgarte-shapiro-2010}. The metric evolution equations do not include matter source terms, since for all the simulations considered in this work we assume that the total mass of the fluid is negligible with respect to the mass of the two BHs, $M_{\mathrm{fluid}} \ll M_{\mathrm{BHs}}$ (i.e., we evolve the Einstein equations in vacuum).

\begin{figure}[!h]
\begin{center} 
\includegraphics[width=.495\textwidth]{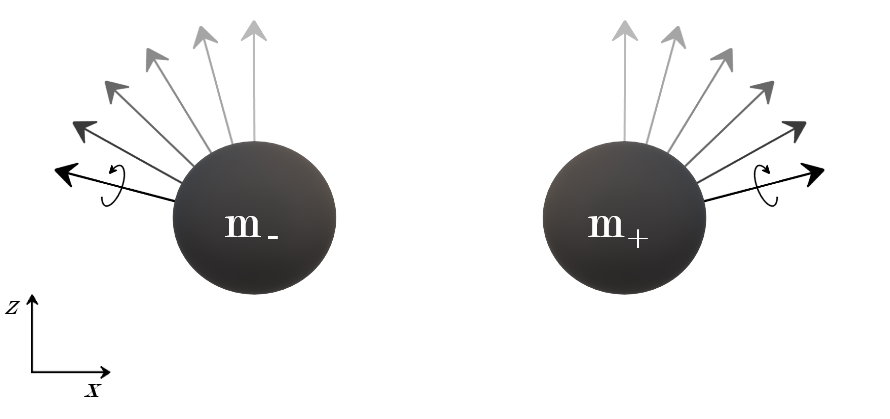}
\end{center}
\caption{Artistic impression of the initial configurations of the binary systems investigated in our simulations. The black holes are initially on the $xy$-plane at an orbital coordinate separation $a_0=12.5M$. The arrows denote the six different orientations of the individual spins, ranging from aligned spins (fainter arrows) to spins nearly orthogonal to the orbital angular momentum (thicker arrows). The initial data for our configurations are given in Table \ref{tab:initdata}.}\label{fig:spins_sketch}
\end{figure}

Our initial metric data are of the Bowen-York type \cite[][]{Bowen-1980}, conditioned to satisfy the constraint equations using the \texttt{TwoPunctures} thorn \cite[][]{Ansorg-2004}. We adopt the standard ``moving puncture'' gauge conditions \cite[][]{Zlochower-2005, Campanelli-2006, vanMeter-2006}:
\begin{equation}
    \partial_t \alpha=-F \alpha^N K
\end{equation}
\begin{equation}
    \partial_t \beta^i= \zeta B^i
\end{equation}
\begin{equation}
    \partial_t B^i=\partial_0 \tilde{\Gamma}^i-\eta B^i
\end{equation}
where we set $F=2$, $N=1$ (1+log condition), $\zeta=3/4$, and $\eta=8/3$.
To generate the initial data we employ the \texttt{NRPyPN} module of \texttt{NRPy+}, a Python-based code generator for numerical relativity \cite[][]{NRPY+, Ruchlin-2018-NRPy+}. In Table \ref{tab:initdata} we list the initial punctures separation, masses, momenta, and spins for our eight binary models. In Fig. \ref{fig:spins_sketch}, we display the six different spin orientations considered in our models. 
\subsection{General relativistic magnetohydrodynamics}\label{sec:mhd_evo}
We evolve the matter and magnetic fields with the \texttt{IllinoisGRMHD} thorn \cite[][]{Noble-2006, Etienne-2015}, which solves the GRMHD equations assuming a perfect fluid stress-energy tensor for the matter in the ideal MHD limit, i.e., considering a perfectly conducting medium. The GRMHD equations are solved in a flux-conservative form via a high-resolution shock-capturing (HRSC) scheme \cite[][]{Etienne-2015}.
The divergence-free property of the magnetic field is guaranteed with the evolution of the magnetic four-vector potential in the ``generalized'' Lorenz gauge introduced in \cite{Farris-2012}. To close the system of GRMHD equations, we must specify an EoS. Across the evolution, we adopt a $\Gamma$-law EOS of the form
\begin{equation}
    P=(\Gamma-1)\rho\epsilon,
\end{equation}
where $\rho$ is the rest-mass density and $\epsilon$ is the specific internal energy.

\begin{table*}[ht]
\centering
\caption{BBH derived quantities in code units of the six $q=0.6$ binary configurations:  merger time $t_{\mathrm{merger}}$, total number of orbits $N_{\mathrm{orb}}$, remnant's mass M$_{\mathrm{rem}}$, components of the remnant's spin parameter $\hat{a}_{\mathrm{rem}}$, remnant's kick velocity $\vec{v}_{\mathrm{kick}}$ and speed $v_{\mathrm{kick}}$ in $\mathrm{km \ s}^{-1}$. Data relative to model b0p2 are missing because of computational issues.}\label{tab:derived}
 \begin{tabular}{lcccccr}
\hline\hline \\[-1.6ex]
  Run & $t_{\mathrm{merger}}$ [M] & \ \  $N_{\mathrm{orb}}$ \ \ &  M$_{\mathrm{rem}}/M$ \ \ &  $\hat{a}_{\mathrm{rem}}$ \ [M$_{\mathrm{rem}}^2$] \ \ &  $\vec{v}_{\mathrm{kick}}$ \ [km \ s$^{-1}$] \ \ & $v_{\mathrm{kick}}$  \ [km \ s$^{-1}$]  \\
  \cmidrule(lr) {1-7}
  \texttt{b0p2}  & 2861  & $\sim$15 &  -  & - &  -  &  -  \\
\cmidrule(lr) {1-7}
 \texttt{b05p12} & 2909  & $\sim$16 &   0.936 &   (0.022, \ 0.005, \ 0.847)  & (35, -877, -1122)  & 1424\\
 \cmidrule(lr) {1-7}
  \texttt{b0p3} &  2958  &  $\sim$16 & 0.938  &    (0.046, \ 0.010, \ 0.819) &    (361, 62, -701) &  791 \\
\cmidrule(lr) {1-7}
  \texttt{b0p4} &  2934  &  $\sim$16 &  0.942  &    (0.068, \ 0.013, \ 0.805) & \ \ (-1027, -155, -381) \ \ &  1107 \\
 \cmidrule(lr) {1-7}
  \texttt{b0p6} &  2829  &  $\sim$15 &  0.946  &    (0.092, \ 0.017,  \ 0.759) &   (636, -337, .135)  & 733 \\
\cmidrule(lr) {1-7}
  \texttt{b0p12}  &  2687  &  $\sim$14  &  0.952 &   (0.113, \ 0.012, \ 0.712) &  (874, 560, 1031)  &  1500 \\
\hline
\hline
 \end{tabular}
\end{table*}
\subsection{Diagnostics}\label{sec:diagnostics}
To follow the dynamics of accretion flows and the evolution of the magnetic fields, as well as their timing features and their relation with the gravitational signal, we employ a number of physical diagnostics.
\subsubsection{Gravitational radiation}
One of the most important outputs of a simulation of merging black holes is the emitted gravitational radiation. A successful approach which can be employed to extract the outgoing GWs in a numerical relativity simulation is based on the calculation of the  Weyl scalar $\Psi_4$ in the Newman-Penrose formalism \cite[][]{Newman-Penrose-1962}, which can be interpreted as a measure of the outgoing gravitational radiation \cite[see, e.g.,][]{Baker-2002} as
\begin{equation}\label{eq:psi4gw}
\Psi_4 = \ddot{h}_+-i\ddot{h}_{\times}. 
\end{equation}
We calculate the Weyl scalar $\Psi_4$ with the thorn \texttt{WeylScal4} \cite[][]{Loffler-2012}. 
Although relation \eqref{eq:psi4gw} is only strictly valid in asymptotically flat spacetimes (i.e., at infinity), in numerical simulations $\Psi_4$ is often extracted on spheres of large but finite radius. In all our simulations, it is extracted on spheres of radius 800 $M$. The GW is then computed integrating Eq. \eqref{eq:psi4gw} in the frequency domain following the standard approach of \citet*{Reisswig-2011}.

\subsubsection{Rest-mass accretion rate}\label{subsec:mdot}
A fundamental diagnostic of our simulations is the flux of rest-mass across the horizon of each BH.
To study the mass accretion rate onto the BH horizons, we use the \texttt{Outflow} thorn \cite[][]{Outflow-thorn}, which computes the flow of rest-mass density across a given spherical surface (e.g., in our case, across each apparent horizon). This quantity is calculated via
\begin{equation}\label{eq:MdotOutflow}
   \dot{M}=-\oint_{S} \sqrt{\gamma}  D v^{i} d \sigma_{i},
\end{equation}
where $D \equiv \rho\alpha u^0$ is the fluid density measured in the observer frame (i.e. $\rho W$, where $W$ is the Lorentz factor), and $\sigma^i$ is the ordinary (flat) space directed area element of the surface enclosing the horizon.

\subsubsection{Frequency analysis}\label{sec:freq_an}
A number of output diagnostics (e.g., the mass accretion rate) extracted from our simulations display quasiperiodic timing features. To examine these periodicities, we perform Fourier analysis on the time-series of the diagnostics (Section \ref{sec:variablity}). The data are prepared according to the following steps: (i) we crop the data excluding from the frequency analysis the first and the last orbit, in order to remove  initial transients and late decrease prior to merger; (ii) we fit the data with a 10th-order polynomial and subtract the fit from the data to isolate the periodic behavior; (iii) we “zero-pad" the data in order to produce a smoother function in the frequency domain. We then proceed to estimate the power spectral density (PSD) with the \texttt{signal.periodogram} method from the \texttt{scipy} package.
\begin{figure*}
\begin{center} 
\includegraphics[width=\textwidth]{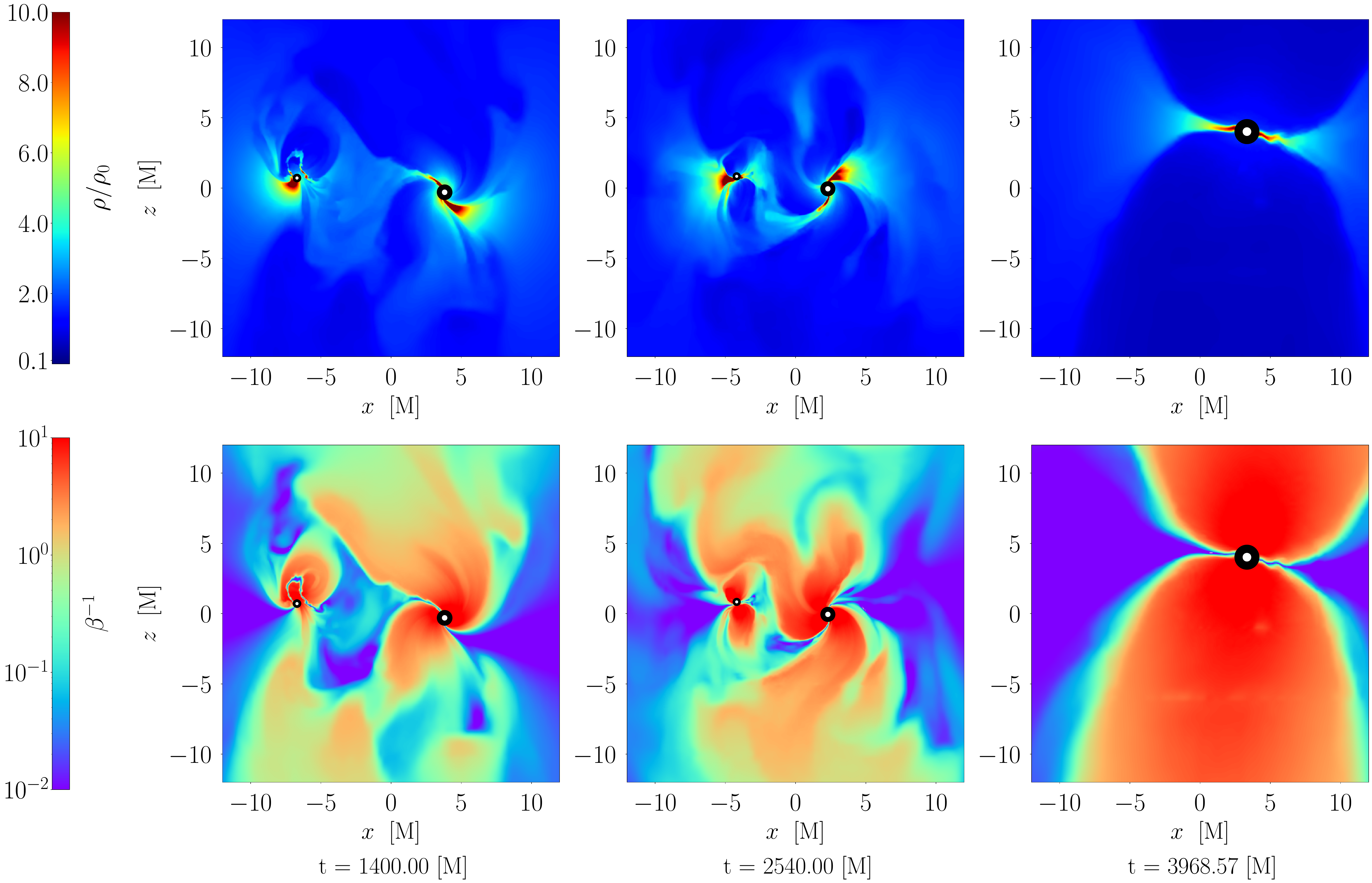}
\end{center}
\caption{Top row: evolution of the normalized rest-mass density $\rho/\rho_0$ on the $xz$-plane of the \texttt{b0p12} configuration. Bottom row: evolution of the magnetic-to-gas pressure $\beta^{-1}$ on the $xz$-plane. Snapshots were taken approximately after five orbits (left), after 11 orbits (middle), and at a time equal to$\sim$1300 M after the merger.} \label{fig:m53V4_panel}
\end{figure*}
\section{Results}\label{sec:results}
In table \ref{tab:derived}, we list a number of derived quantities for our six  spin  configurations All models start at a separation $a_0=12.5 \ M$. We observe that runs \texttt{b0p3} takes the longest time to merge ($t_{\mathrm{merger}}/M=2958$), followed by run \texttt{b0p4} ($t_{\mathrm{merger}}/M=2934$). In general, the merger time is smaller for the configuration with spins (nearly) orthogonal to the orbital angular momentum $L_{\rm orb}$ (run \texttt{b0p12}, $t_{\mathrm{merger}}/M=2861$). Also, run \texttt{b0p12} yields a remnant BH with the smallest spin. Conversely, the remnant mass is largest for run \texttt{b0p12} and becomes smaller as the spin orientation lines up with $L_{\rm orb}$.
The remnant BHs of all misaligned-spin configurations experience a recoil velocity the order of $\sim10^3$ km s$^{-1}$.

\subsection{MHD fields evolution}
The evolution of the rest-mass density for our unequal-mass models exhibits features resembling those by equal-mass configuration \texttt{UUMIS} by \cite{Cattorini-2022}. One noticeable difference lies in the geometry of the disk-like overdensities that develop around the horizons. The plasma around the primary BH settles in wider structures that are tilted about the spin axis. On the top row of Fig.~\ref{fig:m53V4_panel} we display 2D-slices in the $xz$-plane of the rest-mass density $\rho/\rho_0$ for model \texttt{b0p12}. The snapshots were taken approximately after five orbits (left), after 11 orbits (center), and at a time equal to $\sim10^3$ after merger. We observe the development of thin, disk-like structures around the primary BHs, which correspond to the equatorial areas of powerful, magnetically-dominated funnels (protojets) that form near the horizons (bottom rows of Fig.~\ref{fig:m53V4_panel}). The secondary BHs are enveloped by smaller overdensities and develop less powerful protojets. To investigate this feature, we track the evolution of the magnetic field density close to each BH and find that the magnetic field in the vicinity of the primary is on average $\sim$80\% larger than near the secondary. This general trend does not depend on the initial orientation of the individual spins.
\begin{figure*}[ht]
    \centering
        \includegraphics[width=\textwidth]{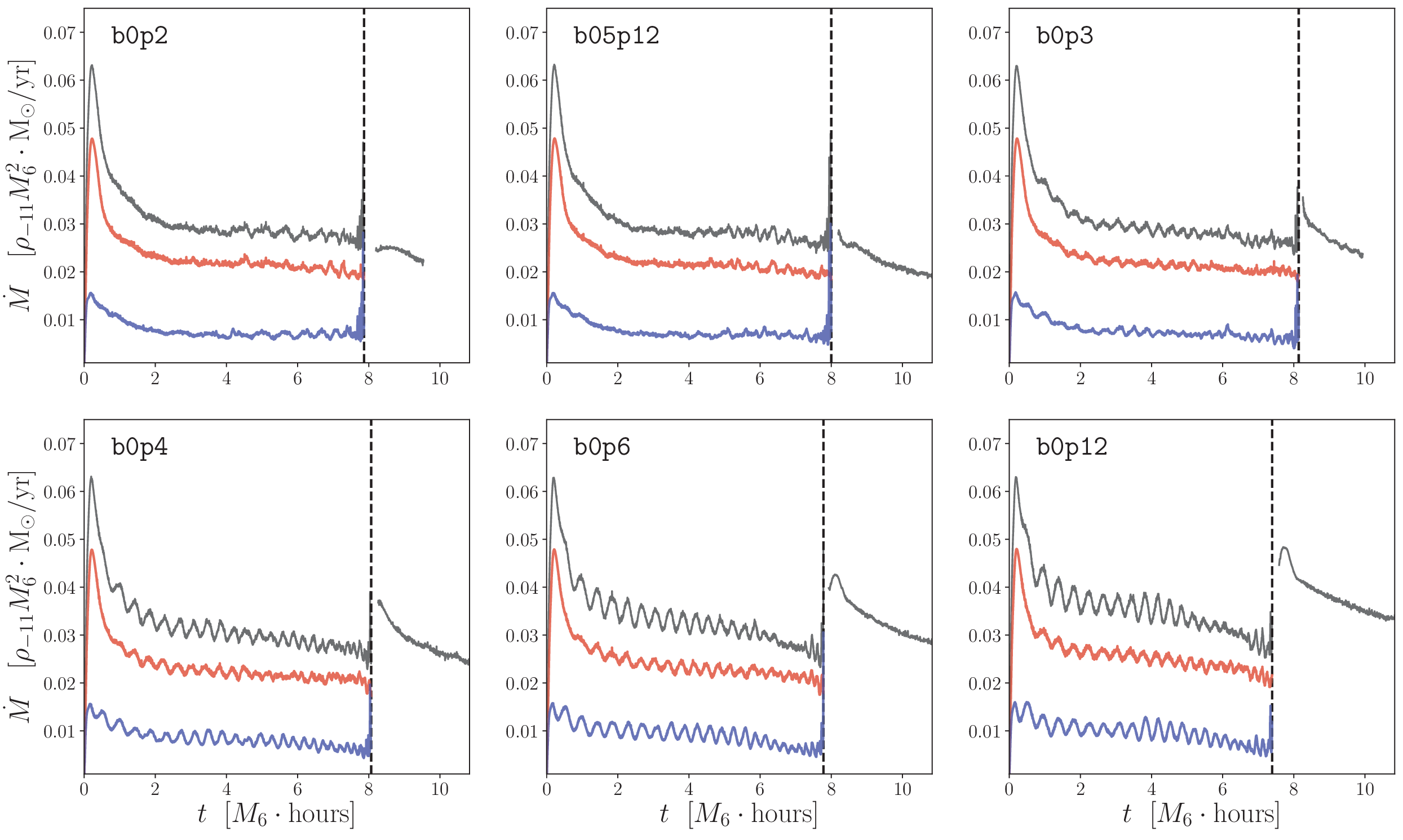}
    \caption{Total rest-mass accretion rate $\dot{M}$ for our six $q=0.6$, $\Gamma=4/3$ models (black lines): the premerger value of $\dot{M}$ is the sum of the individual accretion rates computed on each horizon, the postmerger value is computed on the remnant BH. The colored lines denote the premerger individual rest-mass accretion rates onto the primary (red) and secondary (blue) black holes. The time of merger is marked by black, vertical dashed lines.}\label{fig:m53_all_mdot}
\end{figure*}

The protojets that we observe in the equal-mass simulations from \cite{Cattorini-2022} are in general oriented towards the spin axis of each BH for distances $\lesssim 5 \ M$; for larger separations, they start to align to the orbital axis.
The structure of the protojets of the unequal-mass models is more twisty: for distances $\lesssim 5 \ M$, the funnels are generally oriented towards the BH spin; for separations $5\lesssim d/M \lesssim 50$ their structure is quite convoluted; finally, for distances $\gtrsim50 \ M$, they start to align to the orbital axis. After merger, we observe a magnetic field amplification of $\sim$two orders of magnitude. This is consistent with previous findings by \cite{Giacomazzo-2012, Kelly-2007} and with our results of equal-mass binary models, and appears to indicate a general trend of binary mergers in MMAFs: the magnetic field strength steadily grows as the binary approaches merger, due to the accretion of gas and thus piling up of magnetic field lines in the vicinity of the horizons. 

In the right panels of Fig.~\ref{fig:m53V4_panel}, we observe that the remnant BH is in motion due to a recoil kick velocity. We observe that (i) the remnant is dragging the disk-like structure along, and (ii) the disk-like structure is clearly tilted about the remnant's spin axis. Similarly to the magnetically-dominated funnels emerging from the individual BHs across the inspiral, the protojets emerging from the remnant are directed towards the spin direction for distances $\lesssim 50 \ M$. At larger distances, they start to align toward the $z$-axis.
The recoil kick experienced by massive merged remnants might be associated to a potentially detectable transient, as the BH ``shakes'' the nuclear material and creates shocks which may give rise to an EM signal. 

\subsection{Mass accretion rate}
In Fig. \ref{fig:m53_all_mdot}, we display the total accretion rate $\dot{M}$ for our six $\Gamma=4/3$ models: the premerger value of $\dot{M}$ is the sum of the individual accretion rates computed on each horizon, while the postmerger value is calculated on the remnant BH\footnote{Due to computational issues, the postmerger value of \texttt{b0p2} model in the first moments after the coalescence is missing.}. The time of merger is denoted by a vertical dashed line. The accretion rate is scaled to consider a binary of total mass $M=10^6 \ \mathrm{M}_{\odot}$ (consistent with a binary system in the LISA band) embedded in a gas cloud with initial uniform density $\rho_0=10^{-11}$ g cm$^{-3}$. The units of time are $M_6\cdot$hours, where $M_6\equiv M/10^6\mathrm{M}_{\odot}$.

The time-averaged premerger value of $\dot{M}$ is consistent across all models, but displays a distinct (albeit small) increase as the initial spin orientation shifts from totally-aligned (run \texttt{b0p2}, top-left panel in Fig. \ref{fig:m53_all_mdot}) to nearly orthogonal to the $z$-axis (run \texttt{b0p12}, bottom-right panel).

Noticeably, the oscillatory features of $\dot{M}$ display a more prominent growth: as the initial spins misalign relative to $L_{orb}$, we observe that the variability of the accretion rate exhibits sharper quasiperiodic features; also, the amplitude of the modulations increases. 
In Fig. \ref{fig:m53_all_mdot}, we plot also the individual accretion rates onto the primary (red curve) and secondary (blue curve) black holes. The accretion rate is larger for the primary BH by a factor $\sim$$(m^+/m^-)^2\simeq 2.8$, consistent with Bondi-Hoyle-Lyttleton accretion \cite[see, e.g.][for a review]{Edgar-2004BHL}.
\subsubsection*{Accretion rate variability and spin precession}
We observe that the amplitude of the modulations in $\dot{M}$ appears tightly connected with the inclination of the individual spins. Assuming that the accretion rate translates to detectable accretion-luminosity signals, this finding could in principle  point to a new diagnostic to probe black hole spins observationally. If combined with gravitational wave detections, it could provide an additional means to characterize spin precession, and help distinguishing the timing features that will emerge from the EM candidates to MBBH GW counterparts.

Spin precession is a key feature in the relativistic evolution of two compact objects \cite[][]{Apostolatos-1994} and its measurement has significant impact in both astrophysics and fundamental physics \cite[][]{DeRenzis-2022}. To date, the most commonly used quantity to characterize spin precession in GW data analysis in the so-called \textit{effective precession parameter} $\chi_{\mathrm{p}}$ \cite[][]{Schmidt-2015}. 
\begin{figure}[!h]
\begin{center} 
\includegraphics[width=.495\textwidth]{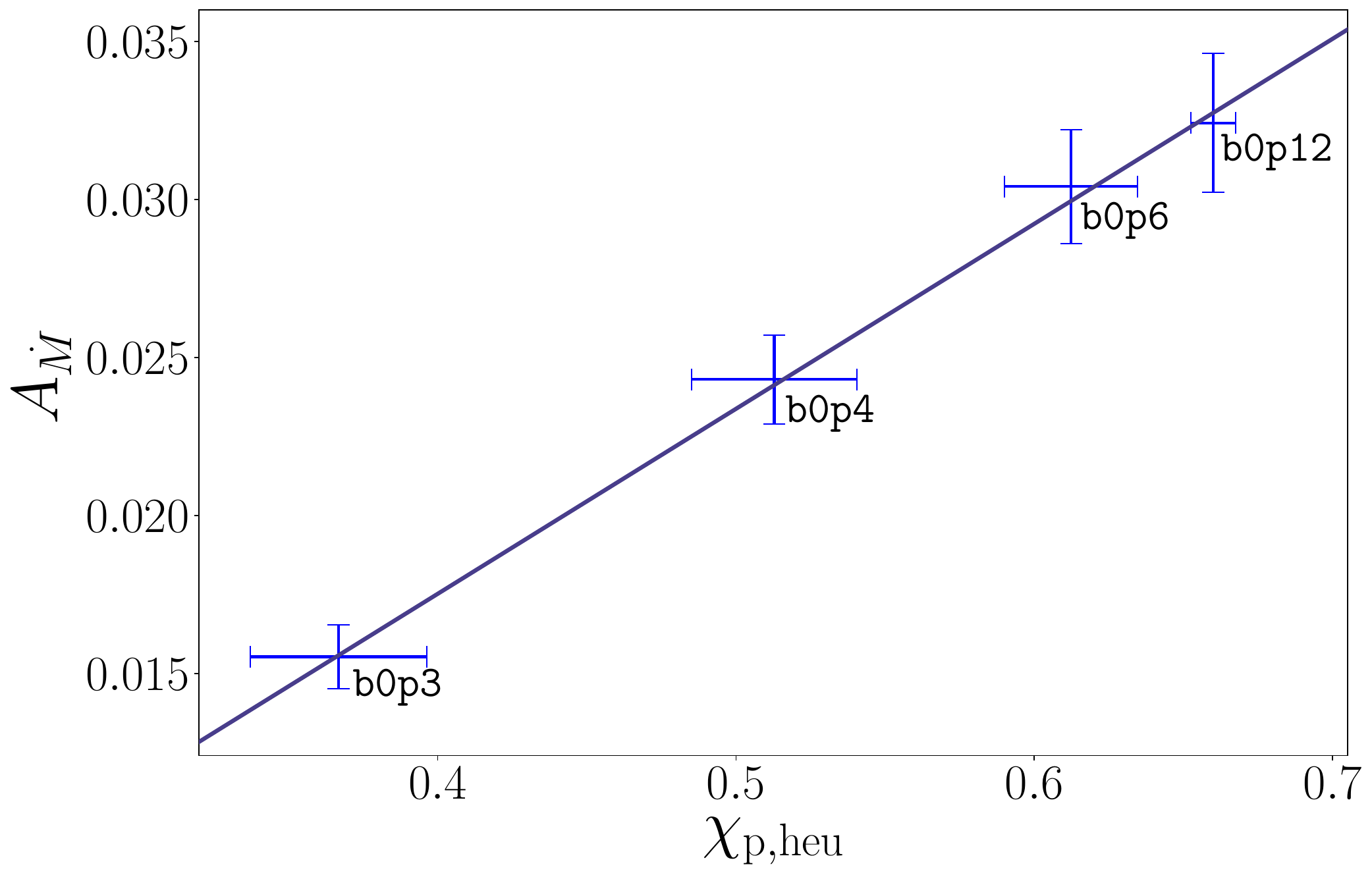}
\end{center}
\caption{Time-averaged amplitudes $\overline{A}_{\dot{M}}$ of the accretion rate as a function of the heuristic spin precession parameter $\chi_{\mathrm{p,heu}}$ for the four $q=0.6,\Gamma=4/3$ models \texttt{b0p3}, \texttt{b0p4}, \texttt{b0p6}, and \texttt{b0p12}. The straight line indicates our best fit. The bars denote the standard normal error.}\label{fig:chi_p_mdot}
\end{figure}
This parameter is widely used in state-of-the-art analyses of GW data to infer the occurrence of precession: a non-zero value of $\chi_{\mathrm{p}}$ is considered a strong indication that orbital precession has been measured \cite[][]{Gerosa-2021b}. We employ the definition of $\chi_{\mathrm{p}}$ given in \cite{Schmidt-2015} (sometimes called “heuristic" spin parameter) and compute the value of $\chi_{\mathrm{p,heu}}$ of four $q=0.6, \Gamma=4/3$ models (i.e., configurations \texttt{b0p3}, \texttt{b0p4}, \texttt{b0p6}, and \texttt{b0p12}). We chose these four models as they display the clearest quasiperiodic pattern in $\dot{M}$. After extracting the average value of $\chi_{\mathrm{p}}$ for these four runs, we wish to connect it to the amplitudes of the modulations in the accretion rate. To this aim, we consider the premerger total accretion rate of the four given models and quantify the strength of the modulations by calculating the \textit{average amplitude} $\overline{A}_{\dot{M}}$ of $\dot{M}$.  To calculate $\overline{A}_{\dot{M}}$, we first crop our $\dot{M}$ data to exclude the early stages of the inspiral and the last phases prior to the merger. This corresponds to considering data in the interval $I=[2,6]M_6$ hours (see Fig. \ref{fig:m53_all_mdot}). Then, we fit the data set with a 10-th order polynomial and subtract it to the accretion rates in order to isolate the oscillatory behavior that we want to investigate.
To calculate $\overline{A}_{\dot{M}}$, we measure the average distance of the data from the mean and normalize it by the corresponding mean value of $\dot{M}$ in the interval $I$.
Then, the values of $\overline{A}_{\dot{M}}$ can be interpreted as percentage-level fluctuations.
In Fig.~\ref{fig:chi_p_mdot}, we plot the values of $\overline{A}_{\dot{M}}$ against the heuristic spin precession parameter $\chi_{\mathrm{p,heu}}$ (the error bars denote the standard normal errors). Remarkably, we observe a clear trend that connects models featuring a stronger spin precession with larger modulations of the accretion rate. 

\subsection{Variability}\label{sec:variablity}
In this section we pursue the investigation of the timing features of $\dot{M}$. In addition, we follow the evolution of the mass $M_{HS}$ enclosed within the Hill spheres of the individual BHs \cite[][]{Paschalidis-2021, Bright-2023}. We investigate periodicities of $\dot{M}$ and $M_{HS}$ calculating the PSD of these quantities for the six $q=0.6, \Gamma=4/3$ models. To account for the diverse features that may arise in unequal-mass systems, we consider separately the accretion rates and the mass enclosed within the Hill sphere of the individual BHs. 

\begin{figure*}
\begin{center} 
\includegraphics[width=.9\textwidth]{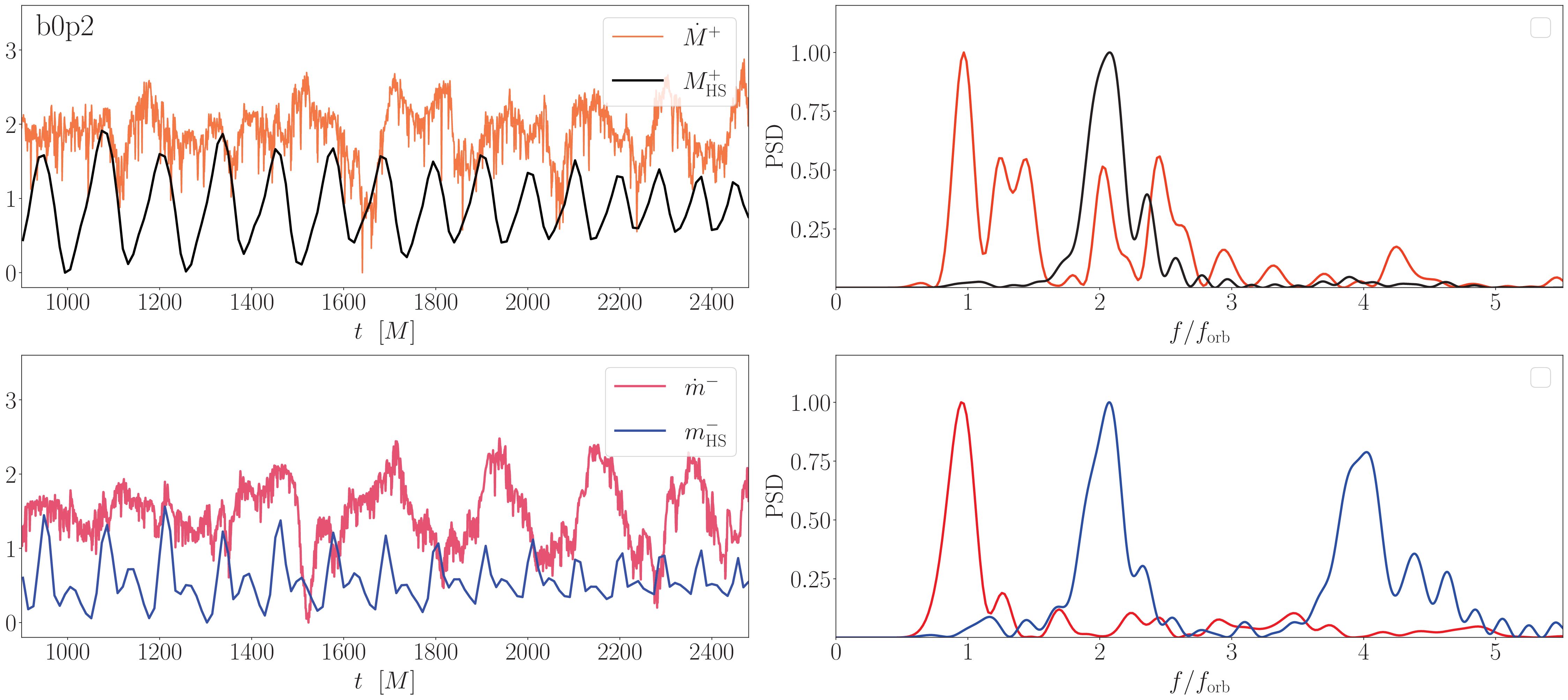}\\
\includegraphics[width=.9\textwidth]{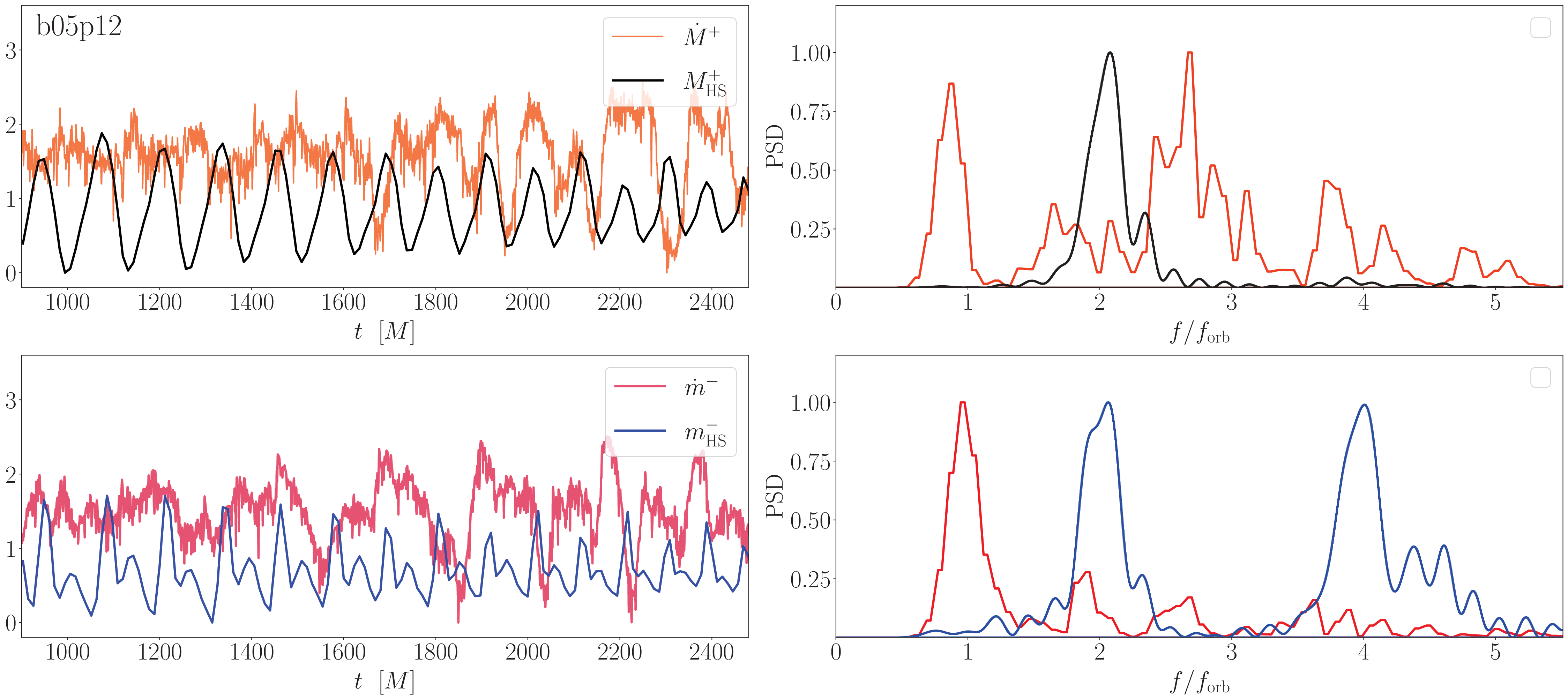}\\
\includegraphics[width=.9\textwidth]{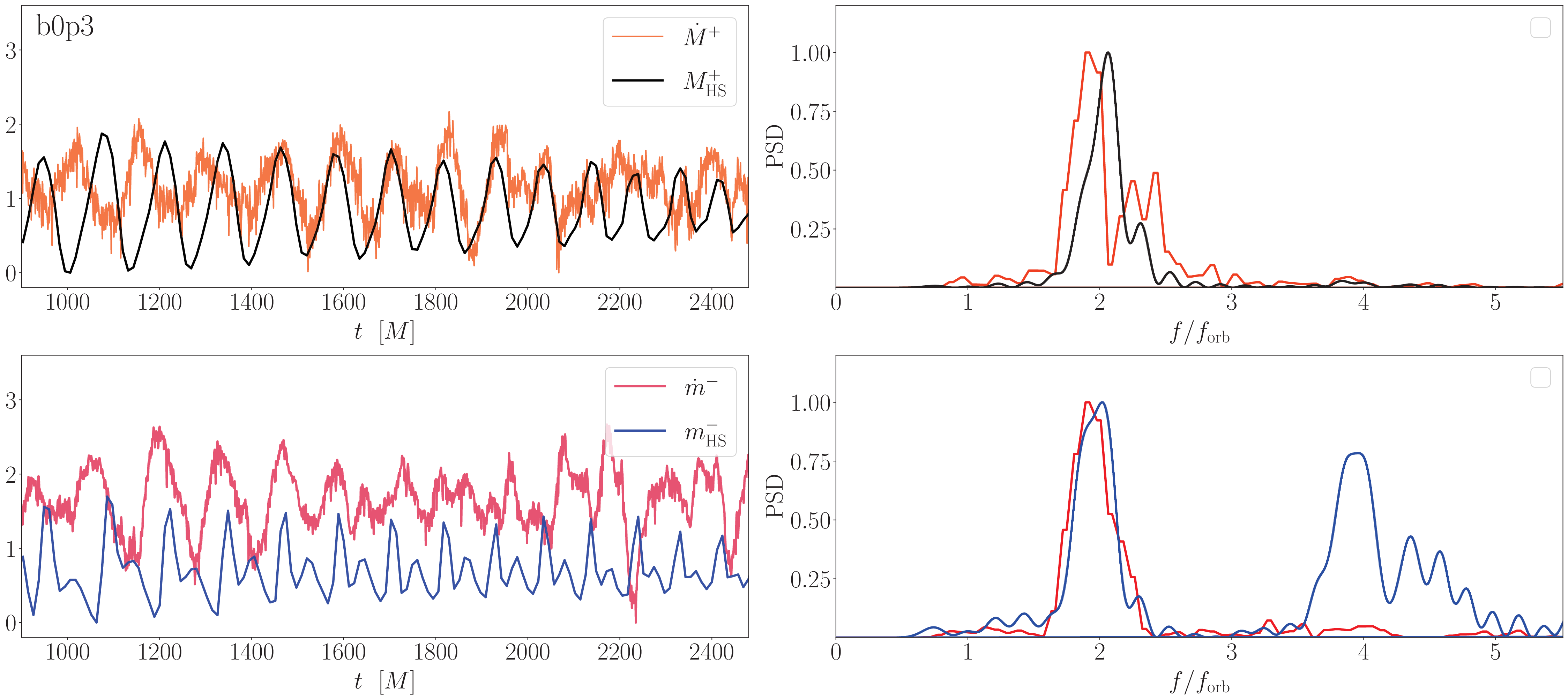}
\end{center}
\caption{Left: Accretion rate and mass contained within the Hill sphere of the primary ($\dot{M}^+ \ M_{HS}^+$) and secondary ($\dot{m}^-, \ m_{HS}^-$) black holes as a function of time for each of our $q=0.6,\Gamma=4/3$ configurations. Bottom: Power spectral densities of the above timeseries normalized by the average binary orbital frequency $f_{orb}$.}\label{fig:mdotHS1}
\end{figure*}

\begin{figure*}
\begin{center} 
\includegraphics[width=.9\textwidth]{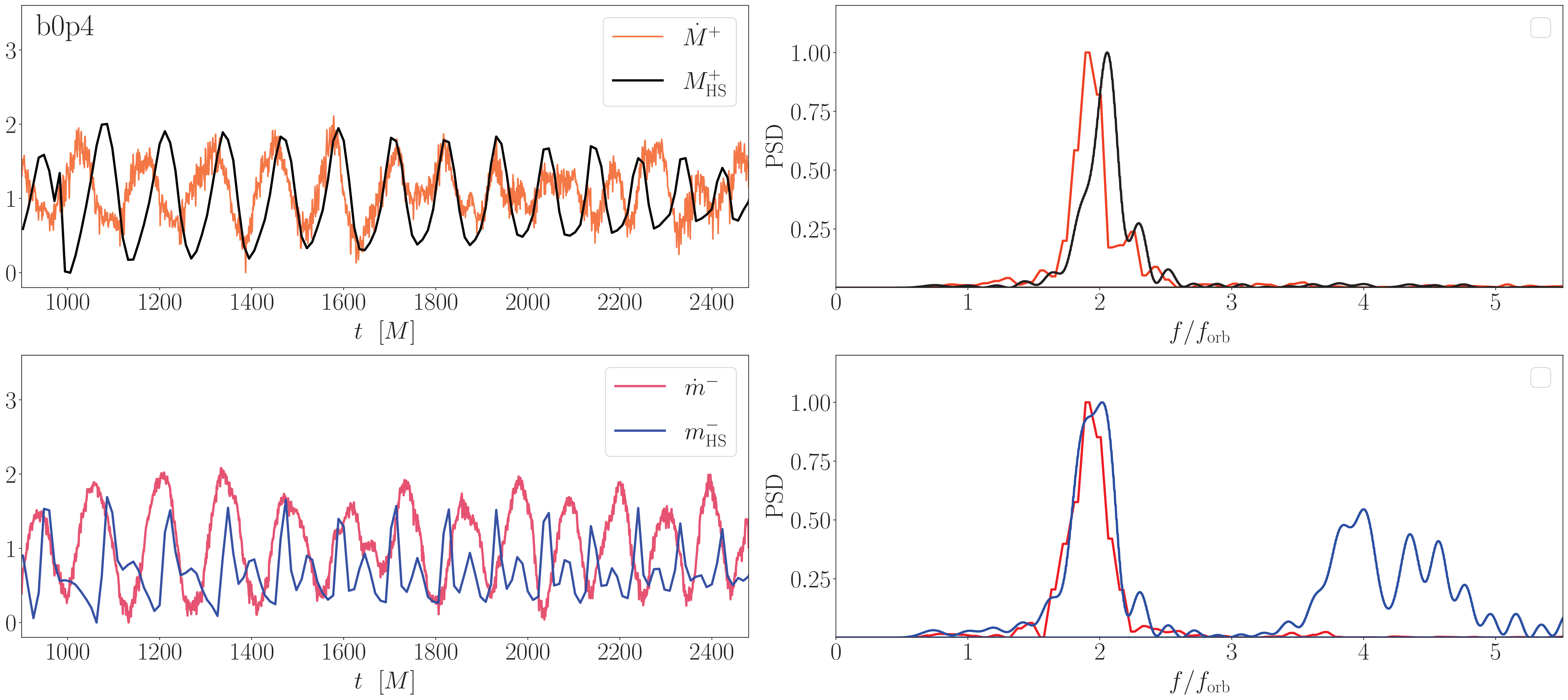}\\
\includegraphics[width=.9\textwidth]{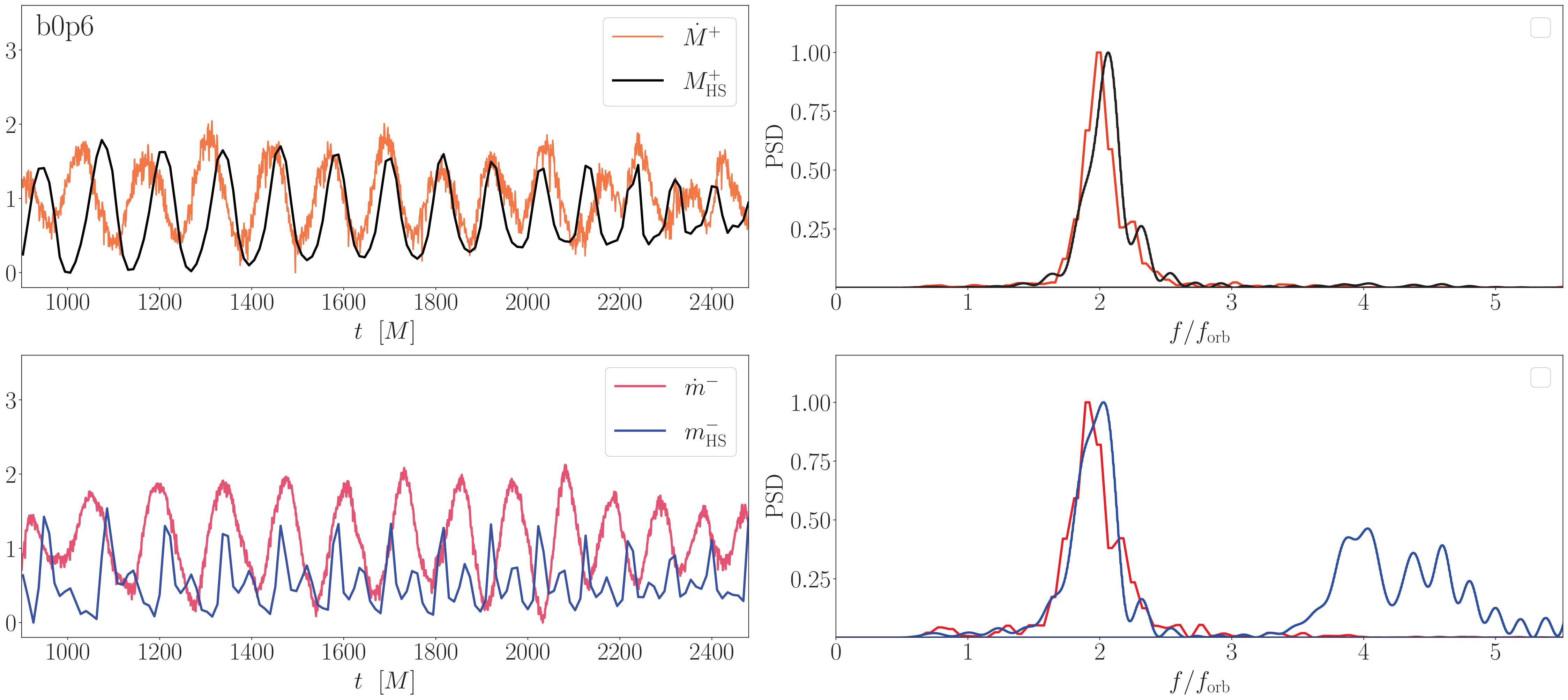}\\
\includegraphics[width=.9\textwidth]{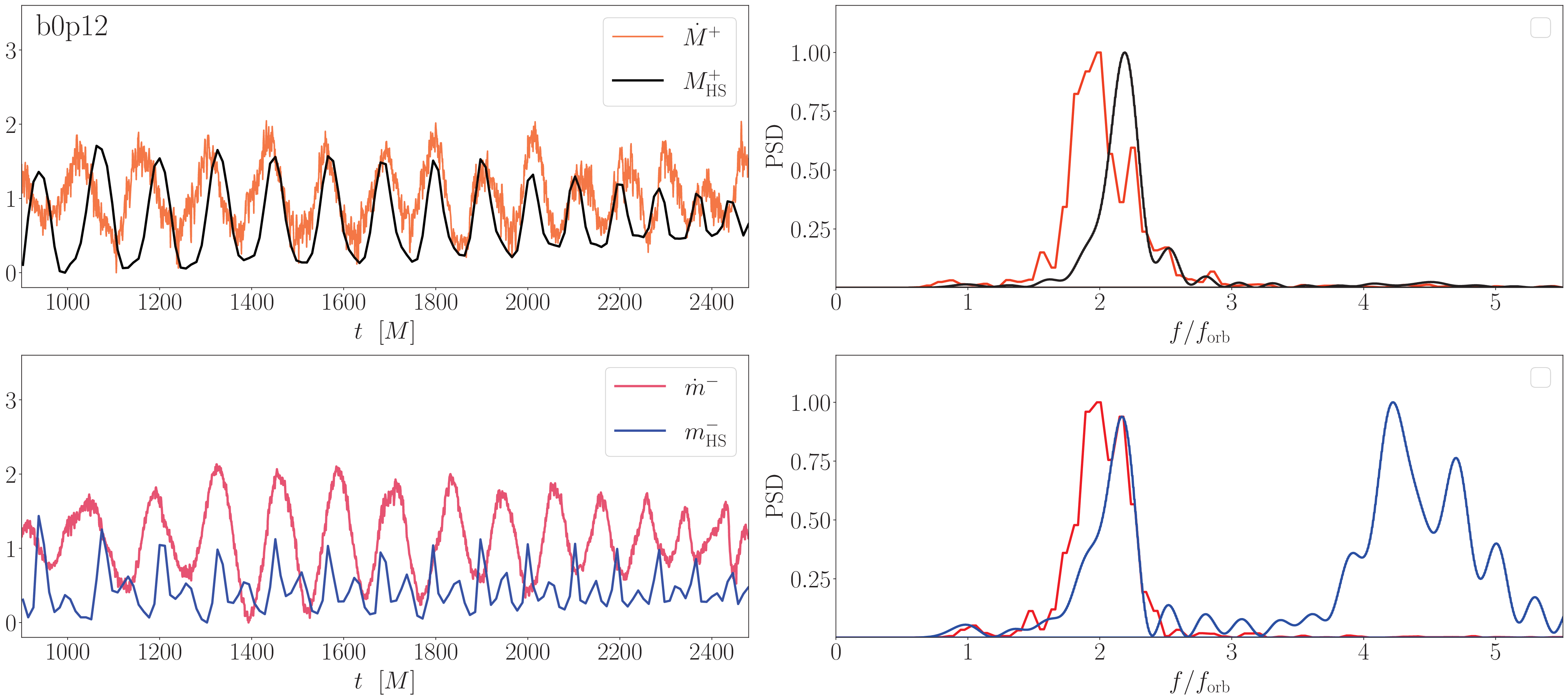}
\end{center}
\caption{(continue) Left: Accretion rate and mass contained within the Hill sphere of the primary ($\dot{M}^+ \ M_{HS}^+$) and secondary ($\dot{m}^-, \ m_{HS}^-$) black holes as a function of time for each of our $q=0.6,\Gamma=4/3$ configurations. Bottom: Power spectral densities of the above timeseries normalized by the average binary orbital frequency $f_{\rm orb}$.}\label{fig:mdotHS2}
\end{figure*}

The left panels of Figs. \ref{fig:mdotHS1}-\ref{fig:mdotHS2} display the accretion rate and mass contained within the Hill sphere of the primary ($\dot{M}^+, \ M_{HS}^+$) and secondary ($\dot{m}^-, \ m_{HS}^-$) black holes as a function of time for each of our $q=0.6,\Gamma=4/3$ configurations; on the right, we show the power spectral densities of the corresponding timeseries normalized by the average binary orbital frequency $f_{\rm orb}$.

The frequency analysis of the mass enclosed within the Hill spheres reveals common features across all spin configurations: the PSD of $M_{HS}^+$ exhibit a prominent peak at $\sim$2-2.2$f_{\rm orb}$; conversely, the PSD of $m_{HS}^-$ show two clear, well-separated peaks falling at $\sim$2-2.2$f_{\rm orb}$ and $\sim$4-4.2$f_{\rm orb}$, respectively. 
The PSD of the mass accretion rates display comparable features for configurations \texttt{b0p3}, \texttt{b0p4}, \texttt{b0p6}, and \texttt{b0p12}. For these models, the accretion rate onto each horizon shows a definite peak at $\sim$2$f_{\rm orb}$. The time evolution of $\dot{M}$ for the remaining configurations (\texttt{b0p2}, and \texttt{b05p12}) is more twisted: for both models, the PSD of the accretion rate onto the secondary BH exhibit a peak at $\sim$$f_{\rm orb}$; instead, the PSD of the accretion rate onto the primary shows in both cases a number of peaks falling between $\sim$$f_{\rm orb}$ and $\sim$4$f_{\rm orb}$.

In general, we find that the modulations in the mass enclosed within the Hill spheres are consistent across all our binary configurations, and display similar features both for the primary and secondary BH. The periodicity in the accretion rate and the mass within the Hill spheres are correlated, and in most cases share the characteristic frequency of $\sim$2$f_{\rm orb}$. This is not the case of models \texttt{b0p2} and \texttt{b05p12}, for which no apparent relation yields between the two quantities.
As pointed out by \cite{Bright-2023}, one could expect a strong correlation between the two quantities, as cycles of increased rest-mass
within the Hill spheres may lead directly to an increase in the mass accretion rate. This picture is consistent with our results for models \texttt{b0p3}-\texttt{b0p12}, but seems in contradiction with the analysis of models \texttt{b0p2} and \texttt{b05p12}.

Such discrepancy will motivate future investigations by our group, which will address a larger number of binary configurations in diverse gaseous environments.

\section{Summary and conclusions}\label{sec:con}
Observed periodicities in the EM light curves rising during the GW chirp might serve as a smoking gun to detect a counterpart of a MBBH merger. Over the last years, it has been suggested that the mass accretion rate onto inspiraling MBBHs can exhibit a quasiperiodic behavior. General relativistic simulations have shown that periodicity signatures in the accretion rate may arise due to a number of mechanisms. In the CBD scenario, streams of gas flung outwards by the binary may periodically hit and shock the inner edge of the disk \cite[][]{Gold-2014b}. Also, modulations due to relativistic Doppler effect are expected to dominate the variability for mass-ration $\leq0.05$ \cite[][]{Haiman-2017}. Modulations in the accretion rate onto MBBHs approaching merger may occur also in a hot and tenuous environment, in which the binary finds itself engulfed in a radiation dominated gas all the way down to the merger \cite[][]{Cattorini-2022}.
Additional variability may be present if the binary is in motion relative to the gas cloud (Bondi-Hoyle-Lyttlelton scenario). If this is the case, modulation in the accretion rate can arise due to shocks that develop around each BH as it moves against the flow of the gas \cite[][]{Farris-2010}.

In this Paper, we presented GRMHD simulations of unequal-mass, misaligned-spinning BBHs and considered a broader family of spin-misalignments to further investigating the link between spin orientation and quasiperiodic features in the accretion rate. We observed that the geometry of the accretion flow is significantly altered by the orientation of the individual spins with respect to the orbital angular momentum $L_{\rm orb}$.
Also, we found that binary systems with a larger spin precession parameter $\chi_{\mathrm{p}}$ imprint a sharper variability in the mass accretion rate. Conversely, individual spins aligned (or nearly aligned) with the orbital angular momentum dampen the strength of the quasiperiodicity.

This result indicates that misaligned spin might imprint periodic variability in the EM light curve of MBBHs approaching the merger.
This would be of particular interest for future space-based gravitational interferometers such as LISA \cite[][]{LISA2023}, allowing a robust identification of the EM counterpart to a GW source.

To extend our study, we aim at exploring a variety of choices for the parameters of the gas (magnetic-to-gas pressure ratio, equation of state) and of the binaries (mass ratio, individual spin magnitude and orientation, eccentricity).
Furthermore, to investigate whether our result is sensitive to the initial gas distribution, a companion Paper \cite[][]{Fedrigo-2023} will focus on the MHD features arising from the interplay of merging BBHs and thinner distributions of gas.
More generally, the in-depth study of MBBHs in diverse gaseous environments is critical in establishing firm predictions of EM signals emerging from these systems; the need for a more robust theoretical characterization of these sources will motivate our future work.

\section*{Acknowledgments}
Numerical calculations were run on the MARCONIA3 cluster at CINECA (Bologna,  Italy) with computational resources provided through a CINECA-INFN agreement (allocation INF22\_teongrav), and on the EAGLE cluster at the Pozna\'{n} Supercomputing and Networking Center (PSNC, Poland) with resources provided by PRACE DECI-17 grant SSMBHB. MC acknowledges support by the 2017-NAZ-0418/PER grant. 

\appendix
\section{Poynting luminosity}\label{appendix:LP}
To get a measure of time dependence of the jet-like Poynting-driven EM power, we compute the Poynting luminosity $L_{\text{Poynt}}$. It is calculated as the surface integral across a two-sphere with radius $R\to\infty$    
\begin{equation}\label{eq:PoyntingLuminosity}
        L_{\mathrm{Poynt}} \approx \lim_{R\to\infty} 2 R^2 \sqrt{\frac{\pi}{3}} S^z_{(1,0)},
    \end{equation}
where $S^z_{(1,0)}$ is the dominant $(l,m) = (1,0)$ spherical harmonic mode of the Poynting vector. In our simulations, we extract the Poynting vector at a distance $R=30 \ M$ and evaluate $L_{\mathrm{Poynt}}$ with Eq.~\eqref{eq:PoyntingLuminosity} \cite[see also][]{Kelly-2017, Cattorini-2021}. 

In our GRMHD simulations, the Poynting luminosity scales as \cite[][]{Cattorini-2021}
\begin{equation}\label{eq:LPscaling1}
    L_{\mathrm{Poynt}} = \rho_0 M^2 F(t/M; \epsilon_0, \zeta_0)
\end{equation}
where $\epsilon_0$ is the initial specific internal energy, $\zeta_0\equiv u_{\mathrm{mag}}/u_{\mathrm{fluid}}$ the initial magnetic-to-fluid energy density ratio and $F$ is a dimensionless function of time (for more details, see Section 3 in \cite{Kelly-2017}).
\begin{figure}[H]
\centering
        \includegraphics[width=.45\textwidth]{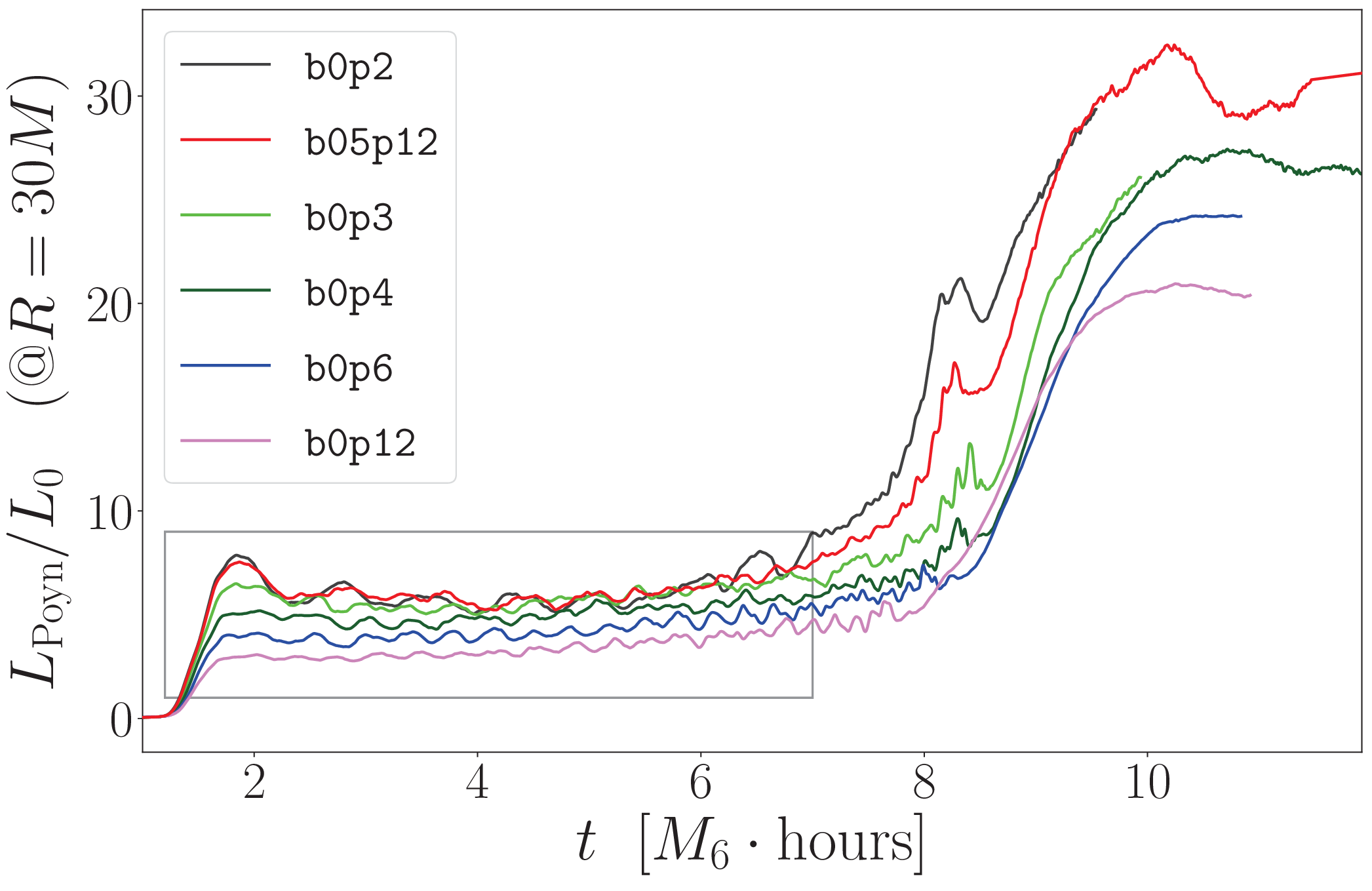}\\
        \includegraphics[width=.45\textwidth]{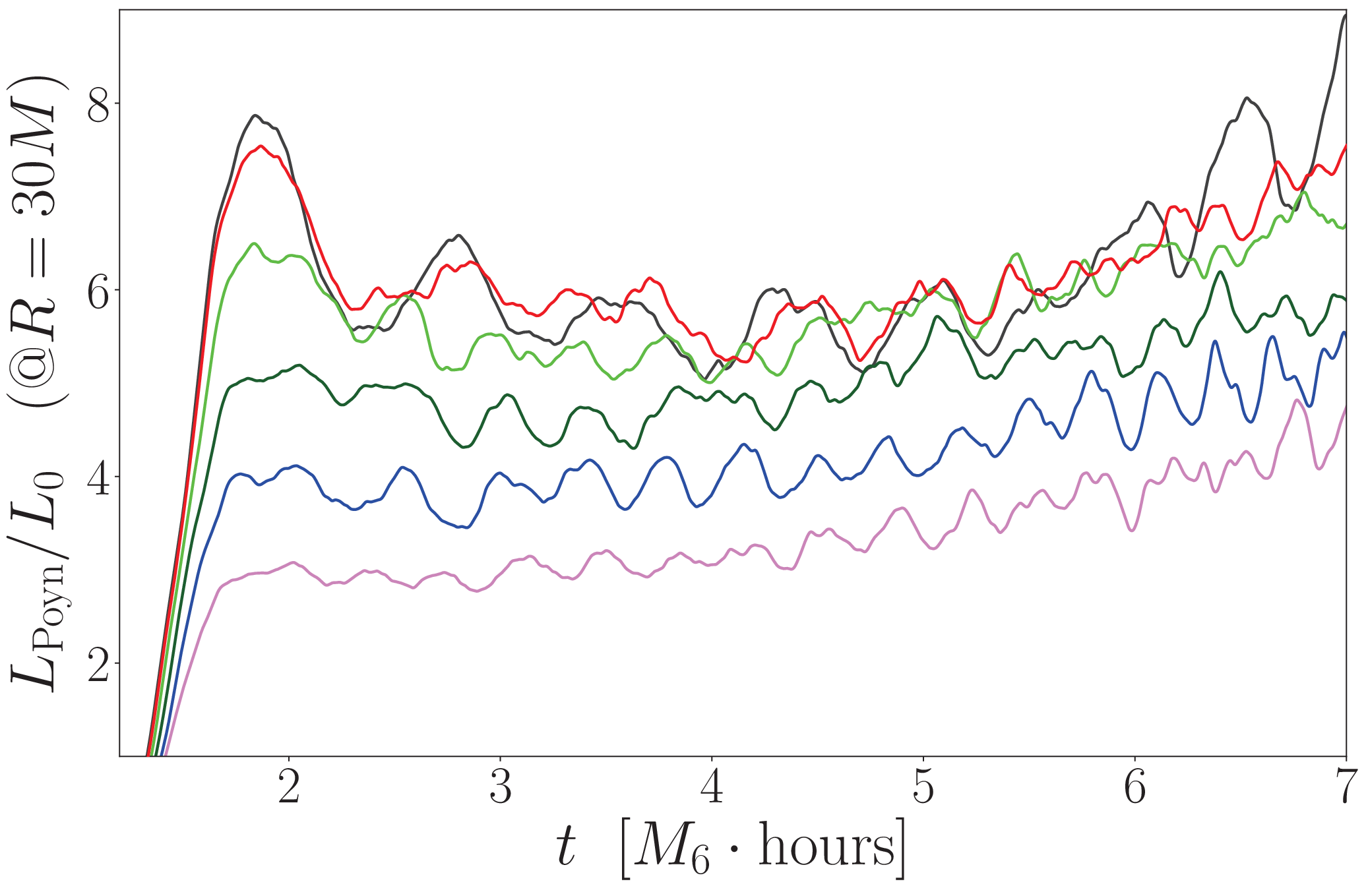}
    \caption{Top: Poynting luminosity $L_{\mathrm{Poyn}}$  extracted on a coordinate sphere of radius $R=30M$ for each $q=0.6,\Gamma=4/3$ model. The values of $L_{\mathrm{Poyn}}$ are normalized to $L_0 \equiv 2.347\times 10^{43} \rho_{-11}M_6^2$ erg s$^{-1}$. Bottom: close-up of the top panel highlighting the features of $L_{\mathrm{Poyn}}$ across the inspiral for different models.}
    \label{fig:m53_LP}
\end{figure}
Equation \eqref{eq:LPscaling1} is in code units, where $c=G=1$. To convert this relation to cgs units, we need to multiply by a factor $G^2/c \approx 1.48 \times 10^{-25}$ g$^{-2}$ cm$^{4}$ s$^{-2}$, and we obtain
\begin{equation}
    \begin{aligned}
L_{\text {Poynt }}(t)=& 1.483 \times 10^{-25}\left(\frac{\rho}{1 \mathrm{~g} \mathrm{~cm}^{-3}}\right)\left(\frac{M}{1 \mathrm{~g}}\right)^{2} \\
& \times F\left(t ; \epsilon_{0}, \zeta_{0}\right) \operatorname{erg} \mathrm{s}^{-1}
\end{aligned}
\end{equation}
If we want to scale with our canonical density $\rho_0=10^{-11}$ g cm$^{-3}$, and for a system of two BHs of $M_1=M_2=10^6$ M$_{\odot}$ (i.e., $M \simeq 3.977 \times 10^{39}$ g), we find
\begin{equation}\label{eq:ourL0}
\begin{split}
    L_{\text {Poynt }}(t)=& \ 2.347 \times 10^{43} \rho_{-11} M_{6}^{2} F\left(t ; \epsilon_{0}, \zeta_{0}\right) \text { erg } \mathrm{s}^{-1} \\
    =& \ L_0 \ \rho_{-11} M_{6}^{2} F\left(t ; \epsilon_{0}, \zeta_{0}\right) \text { erg } \mathrm{s}^{-1}
\end{split}
\end{equation}
where $\rho_{-11}\equiv \rho_0/(10^{-11} \mathrm{g \ cm}^{-3})$ and $M_{6}\equiv M/(10^6 \ \mathrm{M}_{\odot})$.
\begin{figure}[H]
\begin{center} 
\includegraphics[width=.5\textwidth]{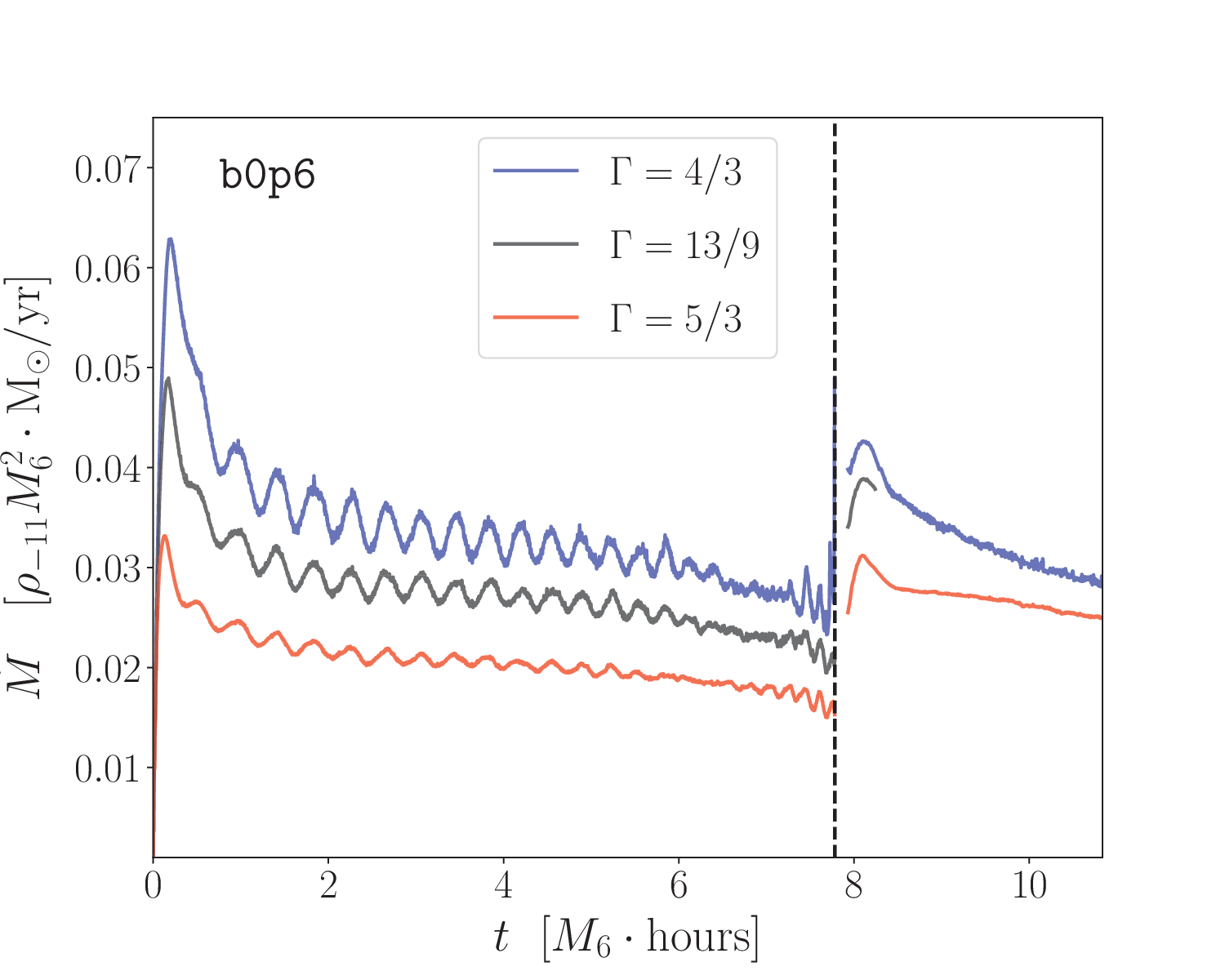}
\end{center}
\caption{Total rest-mass accretion rate $\dot{M}$ for our three \texttt{b0p6} models evolved with different values of the adiabatic index $\Gamma$: the premerger value of $\dot{M}$ is the sum of the individual accretion rates computed on each horizon, the postmerger value is computed on the remnant BH. The merger time is denoted by vertical, dashed lines. The blue (black, red) curve indicates the model evolved with $\Gamma=4/3$ (13/9, 5/3).}\label{fig:mdot_gammas}
\end{figure}
In the following, we examine the features of the EM Poynting emission for our six $q=0.6, \Gamma=4/3$ models.
In Fig. \ref{fig:m53_LP} we display the Poynting luminosity as a function of time extracted on a sphere with radius $R_{\mathrm{ext}}=30 \ M$ (consistently with \cite{Kelly-2017} and our previous works). On the left panel, we show the global evolution of $L_{\mathrm{Poyn}}$ across the full duration of our simulations. On the right panel, we display a close-up that allows to appreciate the features of $L_{\mathrm{Poyn}}$ for different models across the inspiral.

In accordance with our results for equal-mass binaries, we verify that the postmerger peak values of $L_{\mathrm{Poyn}}$ are consistent with the Blandford-Znajek mechanism \cite[][]{BZ-1977} and exhibit a scaling with the square of the spin parameters of the remnants.
During the inspiral, we observe a gradual increase of $L_{\mathrm{Poyn}}$ as the spin of the BHs shifts from being nearly orthogonal to $L_{\rm orb}$ (\texttt{b0p12} model, pink curve on Fig. \ref{fig:m53_LP}) to being aligned with it (\texttt{b0p2} model, black curve).
\section{Dependence on $\Gamma$}
We compute the total rest-mass accretion rate for our three \texttt{b0p6} models that are evolved with different values of the adiabatic index $\Gamma$. By changing $\Gamma$, we can examine accretion flows under a full range of conditions. We consider the values $\Gamma=4/3, 5/3$, indicative of a hot, relativistic gas and a nonrelativistic gas, respectively, and the value $\Gamma=13/9$ as an in-between model. In Fig. \ref{fig:mdot_gammas}, we plot the total rest-mass accretion rate for the three models: we observe  that $\dot{M}$ steadily decreases for higher values of $\Gamma$. This is consistent with the fact that, for fixed values of the rest-mass density, fluids with larger values of $\Gamma$ have a larger pressure that can more effectively balance the gravitational force. Furthermore, we find that a larger adiabatic index yields larger fluctuations in the accretion rate. 

\bibliographystyle{apsrev4-2}
\bibliography{cattorini2023}

\end{document}